\documentclass[aps,prl,floatfox,superscriptaddress,twocolumn]{revtex4-1}
\usepackage{endnotes}
\usepackage{latexsym,amssymb}
\usepackage{amsmath, amsthm, amssymb}
\usepackage{graphicx}
\usepackage{booktabs, longtable, slashbox, multirow}
\usepackage{colortbl}

\pdfoutput=1

\begin{document}
\widetext



\title{Encapsulation by Janus spheroids}

\author{Wei Li}
\email{wel208@lehigh.edu}
\affiliation{Department of Physics, Lehigh University, Bethlehem, PA 18015}

\author{Ya Liu}
\affiliation{Department of Physics, Lehigh University, Bethlehem, PA 18015}

\author{Genevieve Brett}
\affiliation{Department of Physics, Skidmore College, Saratoga Springs, NY 12866}

\author{James D. Gunton}
\affiliation{Department of Physics, Lehigh University, Bethlehem, PA 18015}


\begin{abstract}
The micro/nano encapsulation technology has acquired considerable attention in the fields of drug delivery, biomaterial engineering, and materials science. Based on recent advances in chemical particle synthesis, we propose a primitive model of an encapsulation system produced by the self-assembly of Janus oblate spheroids, particles with oblate spheroidal bodies and two hemi-surfaces coded with dissimilar chemical properties. Using Monte Carlo simulation, we investigate the encapsulation system with spherical particles as encapsulated guests, for different densities. We study the anisotropic effect due to the encapsulating agent's geometric shape and chemical composition on the encapsulation morphology and efficiency. Given the relatively high encapsulation efficiency we find from the simulations, we believe that this method of encapsulation has potential practical value.
\end{abstract}
\maketitle

The subject of encapsulation has received considerable attention in various fields of current research. For instance, in drug delivery, encapsulation with polymer micelles has some remarkable pharmacokinetic advantages, such as reduced systemic toxicity, reduced patient morbidity, and increased bioavailability and absorption efficiency. Encapsulation systems can also be used in specific drug targeting in cancer treatment and controlling the drug release time. This has extensive therapeutic use. \cite{Nature97,PhRe07,PhRe91,IJN08,PPS10,Biom98} In biomaterial research, the encapsulation of proteins could provide the building blocks for the formation of biodegradable scaffolds in tissue engineering. \cite{Biom08,AB10,ABEB06,PR11} Furthermore, a polymer-encapsulating system promises to be industrially important for self-healing material. \cite{AdvM09,Polymer09} Several studies have been devoted to addressing the issue of efficient, uniform encapsulation. Self-directed assembly provides a reasonable method through a bottom-up approach in micro/nanoscale systems. \cite{Sci02G,AdvM02,Sci98,NaMe02,Sci02W} In particular, amphiphilic groups have been used in encapsulation since most encapsulated guests are hydrophobic molecules, that provide a high efficiency, are stable and have the prospect of broad application. \cite{Sci02E,JCP11}

In this article, we present a study of a new encapsulation method by means of the self-assembly of Janus oblate spheroids. Janus particles are two-faced particles, whose chemical makeup differs between the two sides. Due to the recent success in synthesizing these particles, there has been a significant amount of experimental and numerical work reported in recent literatures. \cite{NaMa07,CSR11,JMC08} For instance, Janus spheres have been studied in much of the experimental work, focusing on its synthesis and self-assembly properties. \cite{Sci11,AdMa10,Langmuir08,NaLe06} In addition, some theoretical and numerical studies have also been carried out to investigate the phase diagram and cluster morphology based on simple, primitive, anisotropic potentials. \cite{PRL09,PCCP10,SM11} These studies show an anomalous thermodynamic behavior in the gas-liquid coexistence region, due to the existence of stable micelle and vesicle aggregates in the gas phase (colloidal poor phase). \cite{PRL09} The inner hollow space and the exterior shielding (hydrophilic) surface of the shell structure associated with the micelle or vesicle aggregate makes it a good candidate to act as an encapsulating agent. Furthermore, in the case of Janus oblate spheroids, it has two types of anisotropy; one is due to the Janus character of chemical composition and the other is due to its geometric shape. This reduces the rotational symmetry and also leads to the hollow space inside the shell structure formed by Janus oblate spheroids being larger than in the case of Janus spheres. These effects make it likely that oblate spheroids are more effective encapsulating agents than Janus spheres. Thus in this paper we investigate the performance of encapsulation by Janus oblate spheroids. We believe that Janus oblate spheroids may lead to a high efficiency of encapsulation and serve as a platform for the implantation of functional groups.

\section{Model and Simulation method}
In our model, the encapsulating agents are oblate spheroidal Janus particles, defined as a hemi-spheroid patchy model of oblate spheroids (ellipsoids with the lengths of the principal axes denoted by $c < a = b = \sigma/2$ and aspect ratio $\epsilon = c / a$); the encapsulated guests are chosen to be isotropic hydrophobic spheres with diameter $R = \sigma$. There are three kinds of inter-particle interactions, ellipsoid-ellipsoid ($U_{ee}$), ellipsoid-sphere ($U_{es}$), and sphere-sphere ($U_{ss}$), respectively.

\begin{figure}[htbp] \includegraphics{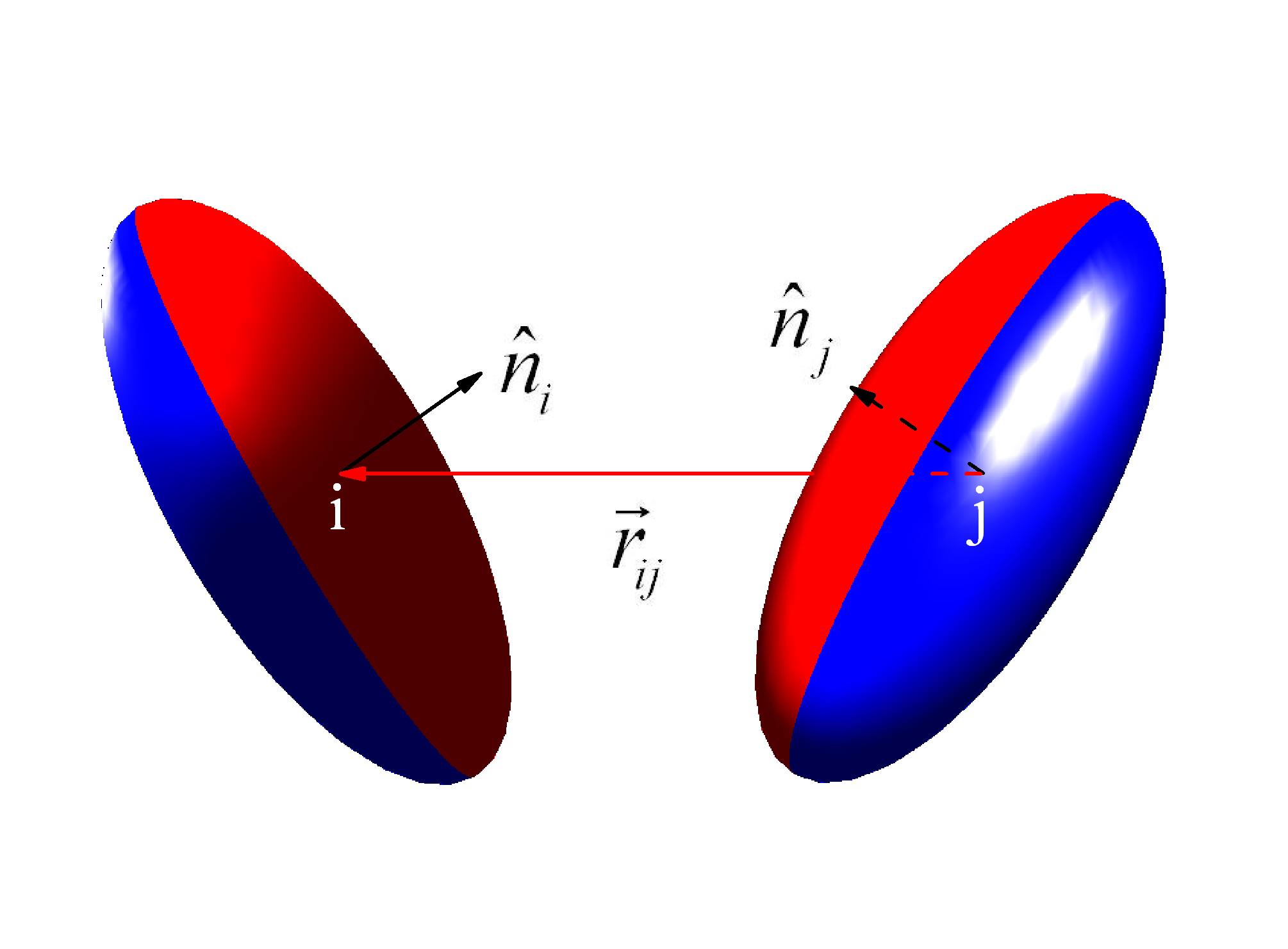}
\caption{(color online). Illustration of Janus oblate spheroids ($\epsilon = 0.4$). The hydrophobic side is coded in red and the hydrophilic side in blue. Red hemi-surfaces attract each other; blue hemi-surfaces interact with other surfaces through a simple hard-core repulsion.}
\label{fig:model}
\end{figure}

Analogous to the patchy spherical model introduced by Kern and Frenkel, we choose the ellipsoids to interact through a pair-potential that depends on their separation and orientation: $U_{ij} = U f(\hat{r}_{ij}, \hat{n}_i, \hat{n}_j)$. \cite{JCP03} The standard square-well potential has been used in earlier work for Janus spheres; however, the determination of the accurate spatial relation between ellipsoids is computationally time-consuming. Thus, as in \cite{Ya11}, we introduce a `quasi-square-well' potential defined as
\begin{eqnarray}
U=
\begin{cases}
\infty& \mbox{if particles overlap}\\
\ -u_0 H(\sigma_{ij}+0.5\sigma - r_{ij})& \mbox{otherwise}\\
\end{cases}
\end{eqnarray}

\begin{eqnarray}
f(\hat{r}_{ij}, \hat{n}_i, \hat{n}_j) =
\begin{cases}
\ 1& \mbox{if } \hat{n}_i \cdot \hat{r}_{ij} \le 0, \hat{n}_j \cdot \hat{r}_{ji} \le 0 \\
\ 0& \mbox{otherwise}
\end{cases}
\end{eqnarray}
where $u_0$ is the well depth, $H(x)$ denotes the Heaviside step function, $r_{ij}$ is the center-to-center distance between particles and $\hat{r}_{ij}$ is the direction of particle separation vector $\vec{r}_{ij}=\vec{r}_{i}-\vec{r}_{j}$, $\hat{n}_{i}, \hat{n}_{j}$ are the patch orientations which are along the semi-minor axes, as shown in Fig. \ref{fig:model}. We treat isotropic hydrophobic sphere as a unique Janus ellipsoid, whereas its $\epsilon = 1.0$, patch covers its whole surface with orientation $\hat{n}_{i} = \hat{r}_{ji} (\hat{n}_{j} = \hat{r}_{ij}$). In this setting, it meets the criterion that the three types of interactions ($U_{ee}$, $U_{es}$, and $U_{ss}$) could be unified in the potential form $U_{ij}$. $\sigma_{ij}$ is introduced as an approximation to characterize the spatial relation from representing molecular interactions by a Gaussian model potential, such that there is no overlapping interaction if $r_{ij} \ge \sigma_{ij}$. \cite{Berne1971} Therefore, in our case $H(\sigma_{ij}+0.5\sigma - r_{ij})$ represents a `quasi-square-well' potential with width $0.5\sigma$. For the hydrophobic interaction between spheres, the potential is a standard square-well potential with well-depth denoted as $u_{ss}$ and a cut-off range $r_c=1.5\sigma$ (since $\sigma_{ij}=\sigma$). To summarize, a) the interactions between two hydrophobic parts (hemi-surfaces of Janus ellipsoids, or entire spherical surfaces) are `quasi-square-well' potentials; b) the interaction between a hydrophilic region and other regions is just a hard-core repulsion.

We use standard Monte Carlo (MC) simulation in the NVT ensemble to study the encapsulation process of this system. We choose small three-dimensional (3D) systems of size $L=24\sigma$ (as the simulation becomes quite computationally expensive with increasing particle number). Periodic boundary conditions are enforced to minimize wall effects. All simulations start from a random initial monomer conformation and the results are averaged over tens of runs. In the following, $\sigma$ is taken as unit of length and $u_0$ the unit of energy. The temperature $T$ is also expressed in the unit of $u_0$ (with the Boltzmann constant $k_B=1$). The other parameters are denoted as follows: the number of ellipsoids $N_e$, the number of spheres $N_s$, the ratio of particle numbers $\alpha=N_e/N_s$, and the total number density $\rho=(N_e+N_s)/L^3$.

\section{Results and Discussion}

Encapsulation is defined such that each guest (sphere) is sufficiently surrounded by agents (oblate spheroids). To be specific, in our model, the self-assembly of the Janus ellipsoids leads to aggregates with one or more outer shells; the outmost surfaces are (mostly) hydrophilic. These shells surround hydrophobic centers consisting of one or more spheres. To study the mechanism of the encapsulation as well as the efficiency dependence on the ellipsoid aspect ratio, we need to find a way to quantify the degree of encapsulation that characterizes how well each sphere is surrounded by ellipsoids. In the first case we study, we choose a low density, $\rho=0.02$ and temperature $T=0.25$, since we find that this thermodynamic point is in the gas phase (micelle) region. We choose a ratio of particle numbers $\alpha=10$, as this equals the number of particles in a stable micelle of Janus spheres \cite{PRL09} and set $u_{ss}=0$ so that there is no hydrophobic attraction between spheres (just hard-core repulsion). We choose $u_{ss}=0$ initially to make it favorable that only one sphere is encapsulated. Later we consider the case for which $u_{ss}\neq 0$. We carry out simulations up to 8 million Monte Carlo steps (MCS). From the results (not shown here), we find a window of simulation time around 2.5 million MCS when mono-dispersed clusters form (with mostly single guests inside), which co-exist with a few monomers and small oligomers. All the simulation results presented in this paper are from the runs in this window, at 2.5 million MCS.

\begin{figure}[htbp] \includegraphics{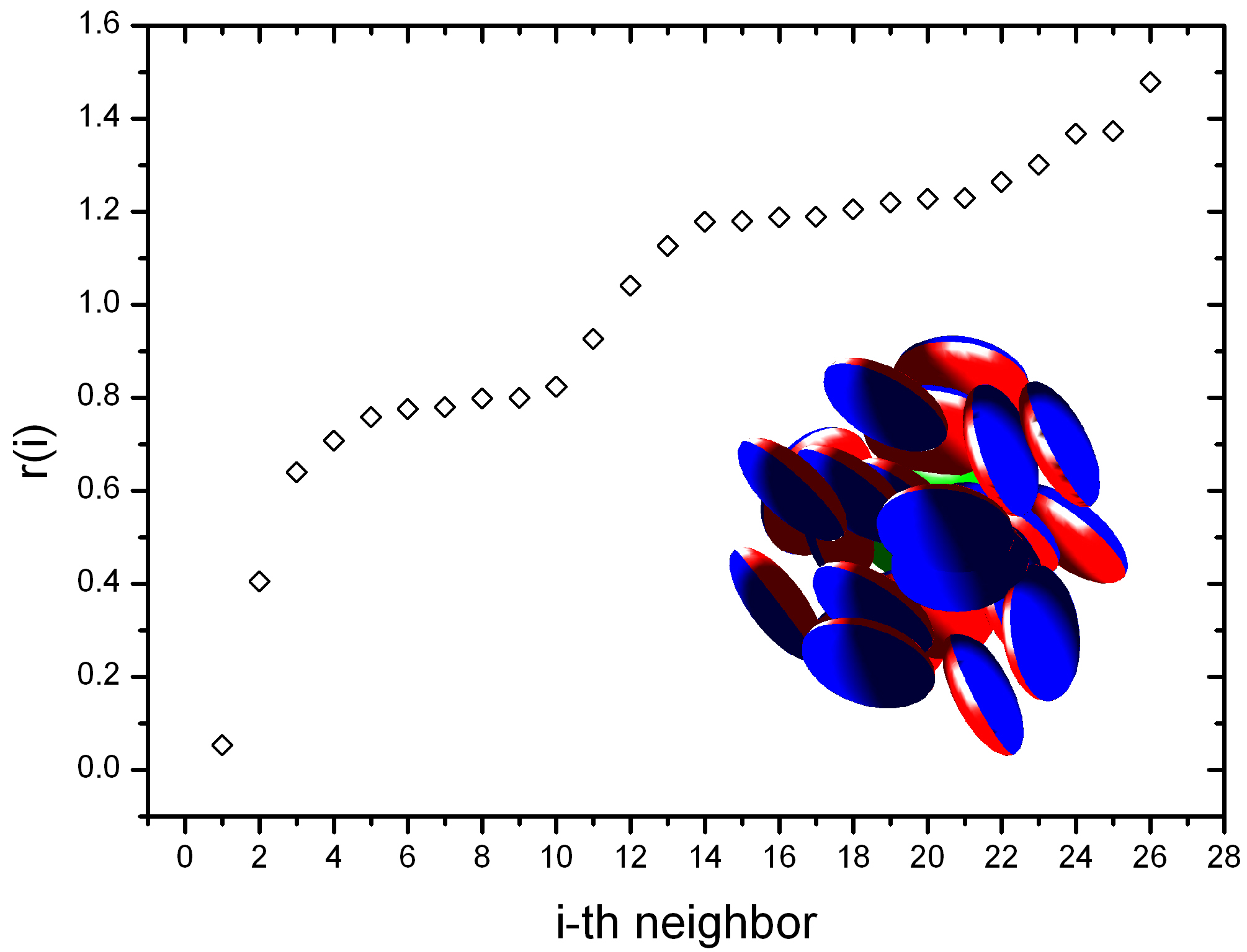}
\caption{(color online). Plot of distance $r(i)$ of the $i^{th}$ particles to the center of mass of the cluster, corresponding to the configuration shown in the inset.}
\label{fig:radial}
\end{figure}

We investigate the clusters formed by plotting the distribution of the particle distance from the cluster center of mass. A typical example is shown in Fig. \ref{fig:radial}. From the result, we find that inside this cluster there are 26 particles with one sphere (colored green) and mostly two layers (shells) of ellipsoids; the inner layer consists of 12 ellipsoids. The number 12 is typical of what we find for the (inner) layer of ellipsoids in other clusters and is also reminiscent of the size of stable Janus sphere micelles. \cite{SM11} Based on this observation, we define the `ideal encapsulation' as corresponding to the case in which each sphere is surrounded by at least 12 ellipsoids so that it forms a cluster with a complete shell. Then we calculate the efficiency $\eta$ by dividing the number of spheres that are `ideally' encapsulated ($N_s^e$) by the number of total spheres inside the system ($N_s$), i.e., $\eta=N_s^e/N_s$.

\begin{figure}[htbp] \includegraphics{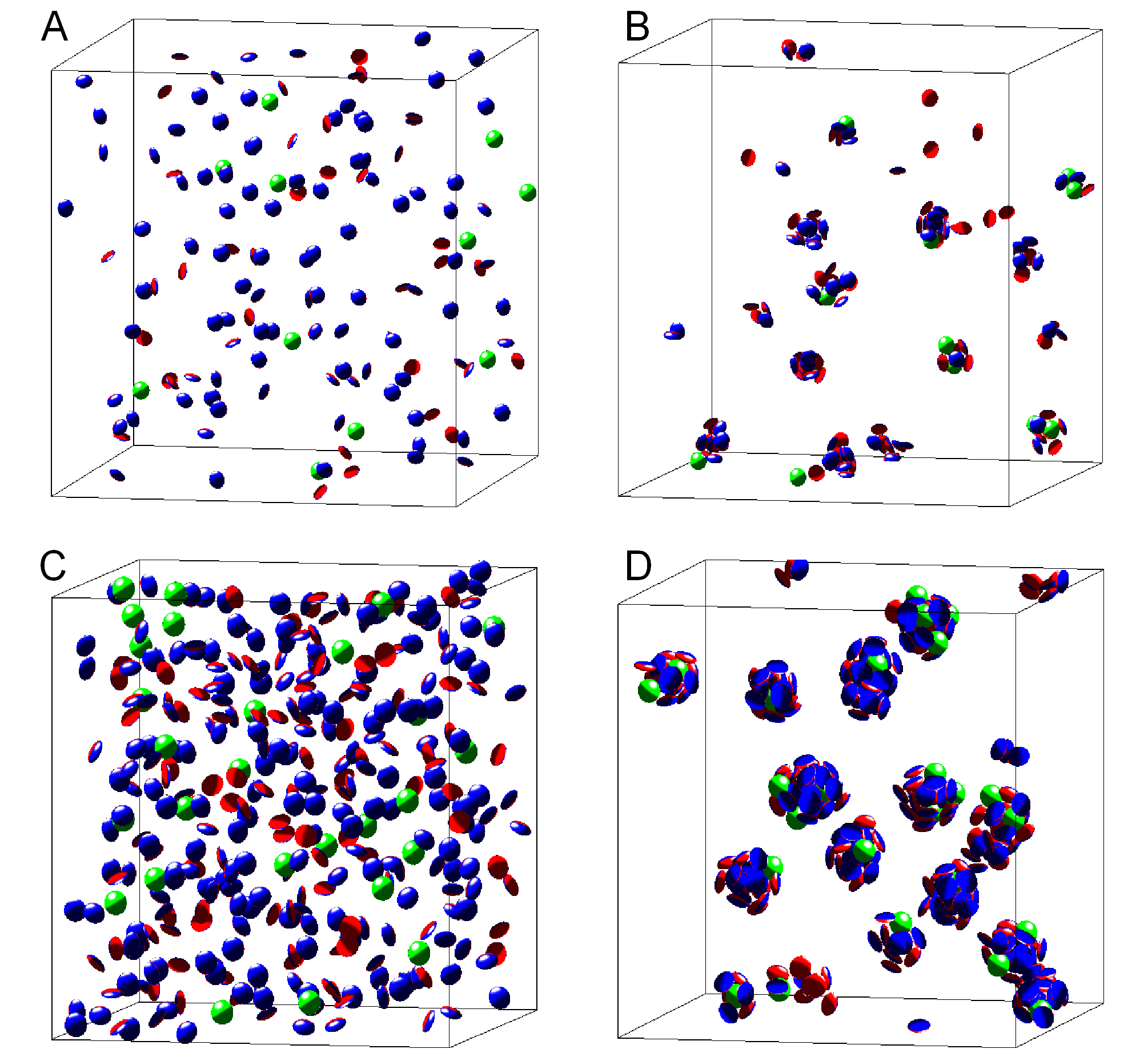}
\caption{(color online). Morphology of the entire system for an ellipsoid aspect ratio $\epsilon=0.4$. (A) Initial configuration of number density $\rho=0.01$. (B) Encapsulation system ($\rho=0.01$) after 2.5 million MCS, efficiency $\eta=9.1\%$. (C) Initial configuration of number density $\rho=0.06$. (D) Encapsulation system ($\rho=0.06$) after 2.5 million MCS, efficiency $\eta=48.1\%$.}
\label{fig:config}
\end{figure}

We choose $\alpha=12, T=0.25$, and run the simulations for different densities and aspect ratios. We show the results in Table \ref{tab:ideal}. The table shows that: a) as the aspect ratio $\epsilon$ varies from 1.0 to 0.3, the efficiency increases and achieves its peak value at $\epsilon=0.6$, then it decreases; b) the same trend is found in the variation with density and reaches another peak at $\rho=0.04$. Two typical runs for density $\rho=0.01$ and $0.06$, with an ellipsoid aspect ratio $\epsilon=0.4$, are presented in Fig. \ref{fig:config}.

\begin{table}
\caption{Efficiency (\%) of `ideal encapsulation' at temperature $T=0.25$}
\centering
\begin{tabular}{l|cccccc}
\toprule[0.8pt]
\backslashbox{$\rho$}{$\epsilon$}&  1.0  &  0.7  &  0.6  &  0.5  &  0.4  &  0.3\\
\hline
0.01 & 15.5 & 9.1 & 13.6 & 10.9 & 7.3 & 11.8 \\
0.02 & 24.7 & 21.3 & 27.3 & 28.0 & 24.0 & 21.3 \\
0.04 & 28.9 & 33.3 & 44.4 & 35.0 & 30.0 & 26.7 \\
0.06 & 23.3 & 26.7 & 25.9 & 28.9 & 32.9 & 40.7 \\
\bottomrule[0.8pt]
\end{tabular}
\label{tab:ideal}
\end{table}

\begin{figure*}[htbp] \includegraphics[width=2\columnwidth]{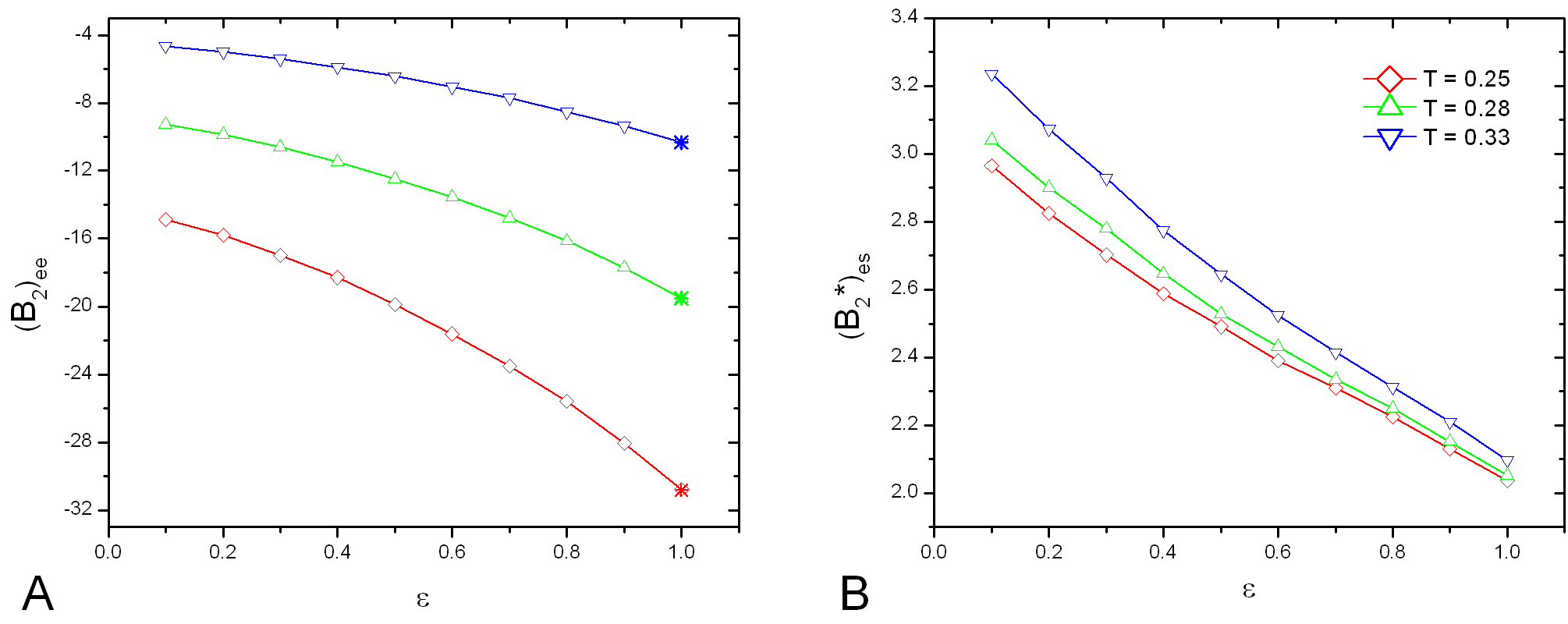}
\caption{(color online). Plot of second virial coefficient ($B_2$) vs. ellipsoid aspect ratio at three temperatures, $T=0.25, 0.28,$ and $0.33$. (A) $B_2$ of Janus ellipsoid-ellipsoid interaction, $(B_2)_{ee}$. Color stars indicate the values from theoretical prediction. (B) Scaled $B_2$ of Janus ellipsoid-sphere interaction on the $(B_2)_{ee}$, i.e. $(B_2^*)_{es}=(B_2)_{es}/(B_2)_{ee}$.}
\label{fig:B2t}
\end{figure*}

The behavior shown in Table \ref{tab:ideal} can be understood in terms of the balance between enthalpy-driven and entropy-driven effects. To illustrate this competition, we first examine the numerical results for the second virial coefficient, $B_2$, that characterizes the pair interaction between the particles, for different aspect ratios at three temperatures. We plot $B_2$ for the Janus ellipsoid-ellipsoid interaction, $(B_2)_{ee}$, and a scaled $B_2$ of the Janus ellipsoid-sphere interaction, with  $(B_2^*)_{es}=(B_2)_{es}/(B_2)_{ee}$, as a function of $\epsilon$ in Fig. \ref{fig:B2t}. The stars in Fig. \ref{fig:B2t} A represent the theoretical prediction for Janus spheres. \cite{JCP03,Ya11} Fig. \ref{fig:B2t} shows a monotonic behavior of these quantities: a) $(B_2)_{ee}$, the magnitude of the effective interactive strength increases as temperature decreases and the aspect ratio goes up. This means that particles combine more strongly at lower temperature; in addition, the larger the aspect ratio, the stronger is the attraction to each other (noticing that with same interaction range, interaction space is proportional to $\epsilon$); b) $(B_2^*)_{es}$ indicates how larger the Janus ellipsoid-sphere interaction is compared to the Janus ellipsoid-ellipsoid interaction. As one would expect, the greater this value is, the more likely an ellipsoid chooses to combine with a sphere instead of another ellipsoid. We see the advantage of a smaller aspect ratio ellipsoid from the curves. This also suggests that such an advantage becomes more important at higher temperatures. However, the absolute values of $B_2$ are relatively small at high temperatures, i.e., the interactions are relatively weak, which is not helpful in forming stable encapsulation structures. From this discussion, we conclude that enthalpy-driven effect should lead to the efficiency increasing as the aspect ratio decreases.

\begin{figure*}[htbp] \includegraphics[width=2\columnwidth]{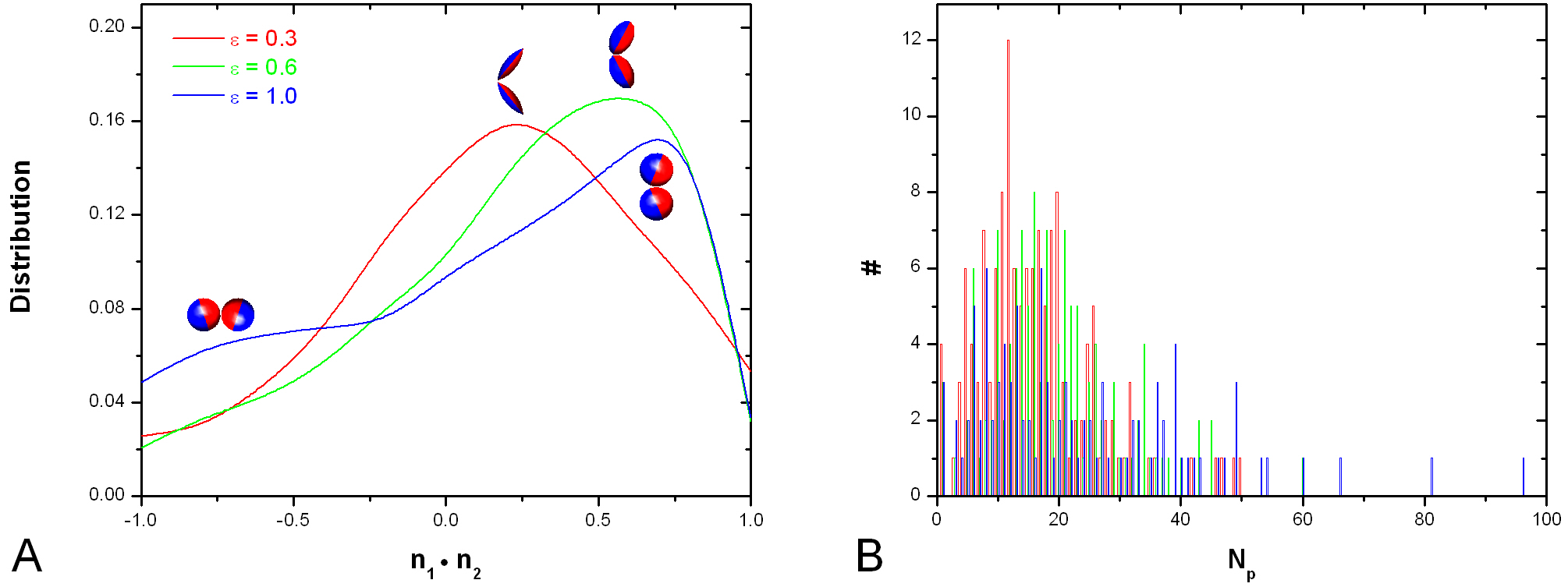}
\caption{(color online). Distribution of (A) orientation correlation over all pairs of bonded ellipsoid particles and (B) cluster size ($N_p$) inside the system after 2.5 million MCS for different ellipsoid aspect ratios, $\epsilon=0.3, 0.6,$ and $1.0$ respectively. }
\label{fig:Np}
\end{figure*}

Next we evaluate the distribution of the scalar product $\hat{n}_1 \cdot \hat{n}_2$ for all pairs of bonded ellipsoid particles in three cases, $\epsilon=0.3, 0.6$ and $1.0$ respectively, and give the results in Fig. \ref{fig:Np} A. We illustrate the relative shapes and orientations of the bonded particles by the conformations shown there. For $\epsilon=1.0$ (Janus spheres), the conformation of pair particles mainly lies in two regimes, facing each other and nearly parallel to each other. This is favorable in forming large clusters with complex structures, providing two types of dissimilar building blocks. For $\epsilon=0.6$ and $0.3$, the curves are localized around single peaks. However, the orientation configuration for $\epsilon=0.3$ is more spread out, which indicates a larger entropy effect due to the freedom of the arrangement of the patch direction. Also, the peak position of $\epsilon=0.3$ around $0.20$ implies that the angle between two patch orientations is $\theta_p=cos^{-1}(0.20)=78.5^\circ$. Therefore, the angle formed between the equatorial planes of the pair of ellipsoids $\theta_e=180^\circ-78.5^\circ=101.5^\circ$. Considering the geometric configuration, the shell structure formed by the pairs with such an internal angle and the other analogous pieces might have a smaller hollow space inside. As a consequence, this is not favored in encapsulation. On the contrary, for $\epsilon=0.6$, the curve is more localized and the peak dominates the distribution, with its pair configuration open enough ($\theta_e=180^\circ-cos^{-1}(0.58)=125.5^\circ$) that the structure aids in the self-assembly of encapsulation. This argument finds additional support in Fig. \ref{fig:Np} B, which shows the cluster size ($N_p$) distribution for three aspect ratios. $\epsilon=1.0$ has a large and extended distribution (up to $N_p=96$), while the histograms of $\epsilon=0.3$ and $0.6$ are much compressed. $\epsilon=0.3$ is peaked sharply around a small size, since it is easier to form a shell structure with a shielding outer surface within small clusters, considering that the pair configurations of $\epsilon=0.3$ have smaller internal angles. Taking the two-dimensional case for example, the equiangular polygon with an internal angle $\theta$ leads to the number of sides $n=360^\circ/(180^\circ-\theta)$. Hence the smaller the internal angle, the fewer sides and fewer vertices the polygon could have. In conclusion, the entropy-driven effect becomes obvious for ellipsoids with small aspect ratios and causes the reduction of encapsulating efficiency. As a consequence, it is reasonable to have $\epsilon=0.6$ as a practical choice for Janus oblate spheroids in encapsulating guests through the self-assembly.

\begin{figure}[htbp] \includegraphics{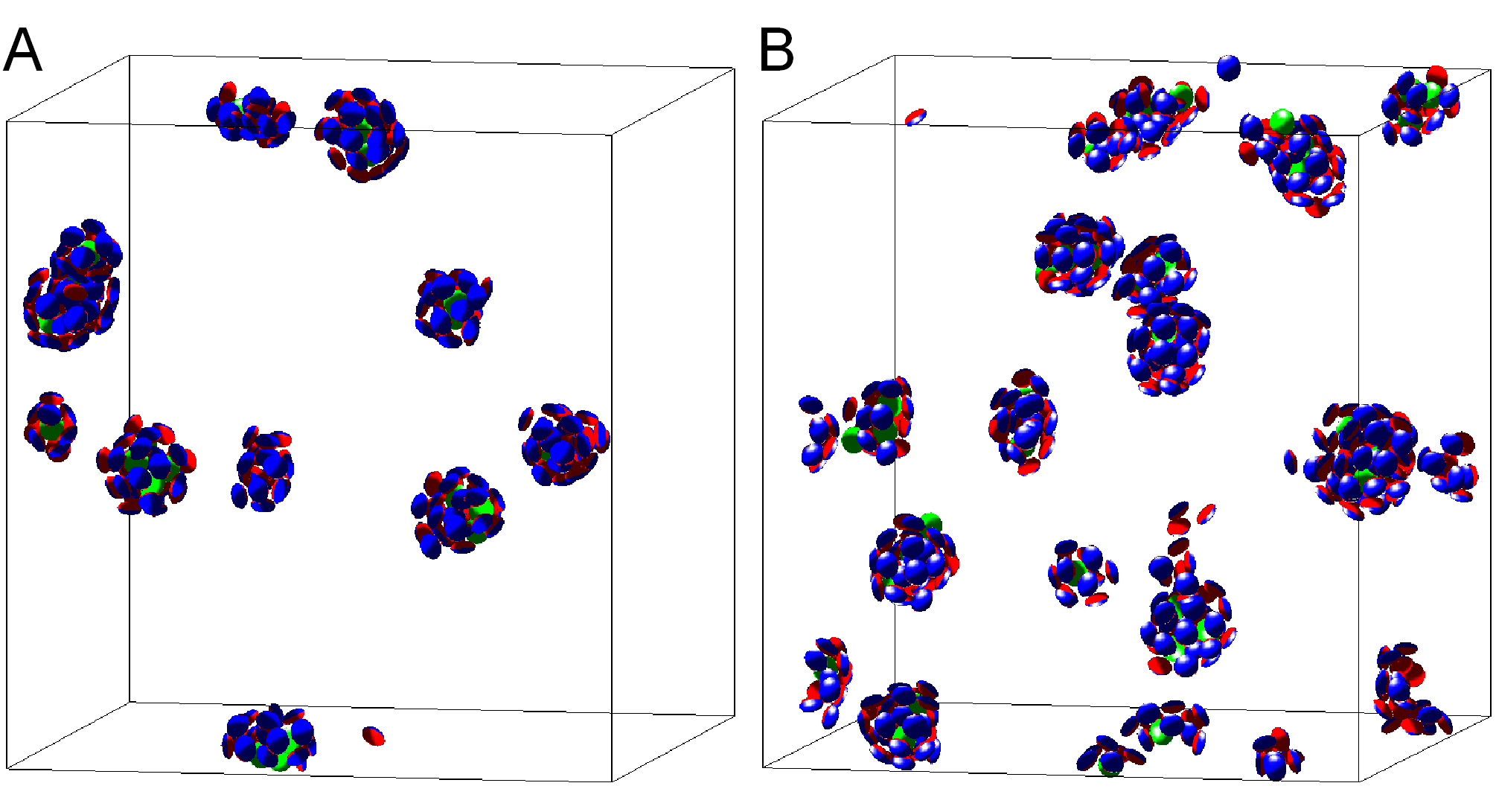}
\caption{(color online). Morphology of entire system for ellipsoid aspect ratio $\epsilon=0.6$ after $2.5$ million MCS. (A) Encapsulation system $\rho=0.02, u_{ss}=u_0$. (B) Encapsulation system $\rho=0.04, u_{ss}=0$.}
\label{fig:config2}
\end{figure}

Now we move on to discuss another factor that might affect the efficiency, i.e., the interaction between encapsulated guests ($u_{ss}$). We start the new runs at $\alpha=12, T=0.25$ for different densities ($\rho=0.02$ and $0.04$) and aspect ratios ($\epsilon=0.5$ and $0.6$) for three values of $u_{ss}$. We consider these more realistic cases in order to compare with practical applications, to shed light on potential experimental work. We introduce a new method to calculate the efficiency, by direct observation from different 3D perspective views. For each run, we view all the clusters generated individually via 3D rotation and count the total number of sphere(s), $n_s$, not sufficiently surrounded to be considered encapsulated. We repeat these observations for each run three times, average the number of spheres `at large' $\bar{n}=\langle n_s \rangle$ and calculate the efficiency $\eta'$ through $(N_s - \bar{n})/N_s$. The results are given in Table \ref{tab:observe}. From these values, we find relatively high efficiencies, which lends credence to the argument that the self-assembly of Janus ellipsoids provides a possible mechanism for efficient encapsulation. Furthermore, we have not found any evidence that the interaction between encapsulated guests would change the encapsulation dramatically at low densities, for the cases we consider here. It is possible that fluctuations and time variation in the cluster equilibration process of different sizes produce the differences in the efficiencies. However, this leaves an open question for the further study of this model at higher densities. To end this section, we show two typical snapshots of the final stages in Fig. \ref{fig:config2} for $\epsilon=0.6$. In both systems, only two guests (spheres) have not been entrapped sufficiently, i.e., only two are not encapsulated. In addition, we have also studied the cases in which there is no interaction between guests and agents, namely $U_{es}=0$; in this case encapsulations were rarely found. The results suggest that by screening the interaction between guests and agents, one could switch off the encapsulation. In this case one would expect to see separate phase separation for each of the two species.

\begin{table}
\caption{Efficiency (\%) of encapsulating from observation at temperature $T=0.25$}
\centering
\begin{tabular}{c | c c c}\toprule[0.8pt]
\backslashbox{$\epsilon$}{$\rho$} & \qquad 0.02 & \qquad 0.04 & \qquad $ u_{ss}/u_0 $\\\midrule[0.5pt]
\multirow{3}{*}{0.5} & \qquad 60.2 & \qquad 62.5 & \qquad 0 \\
& \qquad 73.6 & \qquad 61.4 & \qquad 0.5 \\
& \qquad 65.0 & \qquad 65.9 & \qquad 1.0 \\
\hline
\multirow{3}{*}{0.6} & \qquad 80.7 & \qquad 81.8 & \qquad 0 \\
& \qquad 74.6 & \qquad 63.9 & \qquad 0.5 \\
& \qquad 75.9 & \qquad 81.1 & \qquad 1.0 \\
\bottomrule[0.8pt]
\end{tabular}
\label{tab:observe}
\end{table}

\section*{CONCLUSION}

In summary, we have proposed a potential method for efficient encapsulation of interacting guests, through the self-assembly of Janus oblate spheroids, which based on recent advances in chemical synthesis of Janus particles. We have defined a model correspondingly and used Monte Carlo simulations to investigate the encapsulation process. We find that under the competition of enthalpic and entropic effects, it is most favorable to use Janus oblate spheroids with an aspect ratio $\epsilon=0.6$, at least in the range of ratio values varying from 0.1 to 1.0 (via a step size 0.1). We also examine other possible factors in determining the efficiency, such as density, temperature, and interaction strength etc. The results turn out to support the argument that a method based on anisotropic properties introduced by geometric shape and chemical composition could be quite general in its application. Thus we believe the methodology of encapsulation we propose here could be of practical value and raise questions for future study.

\section{Acknowledgements}
This work was supported by a grant from the Mathers Foundation. One of us (GB) was supported by the NSF REU Site Grant in Physics at Lehigh University. We thank Wenping Wang at The University of Hong Kong for providing the ellipsoid code.

\end{document}